%% file: FromNewtonToCellularAutomata.tex
\documentclass[a4paper,10pt]{article}
\usepackage[latin1]{inputenc}
\usepackage{color}
\usepackage[dvips]{graphicx}
\usepackage{amsmath}
\usepackage{hyperref}
\usepackage[margin=2cm]{geometry}
\definecolor{refkey}{rgb}{0,0,1}
\definecolor{labelkey}{rgb}{0,0,1}
\include{macros}
\usepackage{rotating}
\date{\today}
\title{From Newton to Cellular Automata}
\author{Franco Bagnoli\\
Dept. Energy and CSDC\\
University of Florence\\Florence, Italy\\
and INFN, sec. Firenze\\
franco.bagnoli@unifi.it
}

\begin{document}
\maketitle
\tableofcontents

\begin{abstract}
I outline a possible logical path from the formulation of physics of classical mechanics to
``abstract'' systems like cellular automata. The goal of this article is that of illustrating
why physicists often study extremely simplified models, instead of just numerically
integrating the fundamental equations of physics. This exposition is obviously only partial and 
based on my expertise and my interests. 

A similar version of this text appeared under the title \emph{Interaction Based Computing in Physics} in the \emph{Encyclopedia of Complexity and System Science},
Springer, New York 2009 p.\ 4902.
\end{abstract}

\section{Introduction}
Physics investigation is based on building models of reality: in
  order for a phenomenon to be \emph{understood}, we need to
represent it in our minds using a limited amount of symbols. However,
it is a common experience that, even using simple
 ``building blocks'' one usually obtains systems whose behavior is
quite complex. In this case one needs to develop new languages and
new \emph{phenomenological} models in order to manage this
``complexity''.

Computers have changed the way a physical model
is studied. Computers may be used to \emph{calculate} the
properties of a very complicated model representing a real system, or
to investigate \emph{experimentally} what are the essential
ingredients of a complex
phenomenon. In order to carry out these explorations, several basic
models have been developed, which are now used as building blocks for
performing simulations and designing algorithms in many fields, from
chemistry to engineering, from natural sciences to psychology. Rather
than being derived from some fundamental law of physics, these blocks
constitute \emph{artificial worlds} still to be completely explored.

In this article we shall first present a pathway from Newton's laws to
cellular automata and agent-based simulations, showing (some)
computational approaches in
classical physics. Then, we shall present some examples of
\emph{artificial worlds} in physics. 

\Section[Intro]{Physics and computers}

Some sixty years ago, shortly after the end of Second World
War, computers become available to
scientists. Computers were
used during the last years of
the war for performing computations about the atomic
bomb~\cite{HistoryComputing,LosAlamosComputing}. 

Up to then, the only available tool, except experiments, was paper and
pencil. Starting
with Newton and Leibnitz,  humans discovered that continuous
mathematics (\ie, differential and integral calculus) allowed
to derive many consequences of the a given hypothesis just by
the manipulation of symbols. It seemed natural to express all
quantities (\eg, time, space, mass) as continuous
variables. Notice however that the idea of a continuous number is
not at all ``natural'': one has to learn how to deal with it, while
(small) integer numbers can be used and manipulated (added,
subtracted) by illiterate humans and also by many animals. A point
which is worth to be stressed is that any computation refers to a
model of certain aspects of reality, considered
most important, while others are assumed to be not important

Unfortunately most of human-accessible explorations in
physics are limited to almost-linear systems, or systems whose
effective number of variables is quite small. On the other hand, most
of naturally occurring phenomena can be ``successfully'' modeled only
using nonlinear elements. Therefore, most of pre-computer physics is
essentially linear physics, although 
astronomers (like other scientists) used to integrate numerically, by
hand, the non-linear equations of gravitation, in order to compute
the trajectories of planets. This computation, however, was so
cumbersome that no ``playing'' with trajectories was possible.

While analog computers have been used for integrating
differential equations, the much more flexible digital computers
are deterministic discrete systems. The
way of working  of a (serial) computer is that of a very fast
automaton, that manipulates data following a program. 

In order to use computers as fast calculators, 
scientists ported and adapted existing numerical algorithms, and
developed new ones. This implied the development of
techniques able to \emph{approximate} the computations of continuous
mathematics using computer algebra. However, numbers in computers are
not exactly the same as human numbers, in
particular they have finite (and varying) precision. 
 
This intrinsic precision limit has deep consequences in the
simulations of nonlinear systems, in particular of chaotic ones.
Indeed, chaos was ``numerically discovered'' by
Lorenz~\cite{LorenzChaos} after the observation that a simple
approximation,  a number that was retyped with fewer decimals, caused
a macroscopic change in the trajectory under study.

With all their limits, computers can be fruitfully used \emph{just} to
speed-up computations that \emph{could} eventually be performed by
humans. However, since the increase in velocity is of several order of
magnitude, it becomes possible to include more and more details into
the basic model of the phenomenon under investigation, well beyond
what would be possible with an army of ``human computers''. The idea
of exploiting the \emph{brute power} of fast computers has originated
a fruitful line of investigation in \emph{numerical physics}
especially in the field of chemistry, biological molecules, structure
of matter. The power of computers has allowed for instance to include
quantum mechanical effects in the computation of the structure of
biomolecules~\cite{CarParrinello}, and although these techniques may
be targeted  as ``brute force'', the algorithms developed are actually
quite sophisticated.

However, a completely different usage of computers is possible:
instead of exploiting them for performing computations on models that
already proved to approximate the reality, one can use computers as
``experimental'' apparatus to investigate the patterns of
\emph{theoretical} models, generally non-linear. This is what Lorenz
did after having found the first examample of chaos in computational
physics. He started simplifying his equations in order to enucleate
the minimal ingredients of what would be called the \emph{butterfly
effect}. 

Much earlier than Lorenz, Fermi, Pasta and Ulam (and the
programmer Tsingou~\cite{FPUPedagogical}) used one on the very first
available computers to
investigate the basis of statistical mechanics: how energy
distributes among the oscillation modes of a chain of nonlinear
oscillators~\cite{FPUOriginal}. 

Also in this case the model is simplified at its maximum, in
order to put into evidence what are the fundamental ingredients of
the observed pattern, and also to use all the available power of
computers to increase the precision, the duration and the size of the
simulation.

This simplification is even more fruitful in the study of
systems with many degrees of freedom, that we may denote
generically as \emph{extended systems}.
We humans are not prepared to
manipulate more that a few symbols at once. So, unless
there is a way
of grouping together many parts (using averages, like for instance
when considering  the pressure of a gas as an average over
extremely many particle collisions), we are in difficulties in
\emph{understanding} such
systems.
They may
nevertheless be studied performing ``experiments'' on computers. 
Again, the idea is that of simplifying at most the original model, in
order to isolate the fundamental ingredients of the observed
behavior. It is therefore natural to explore systems whose
\emph{physics} is different from the
usual one. These \emph{artificial
worlds} are preferably formulated in discrete terms, more suitable to
be implemented in computers (see \secref{ArtificialWorlds}).

This line of investigation is of growing interest today: since modern
computers can easily simulate thousands or millions of
\emph{elementary
automata} (often called \emph{agents}), it may be possible to design
\emph{artificial worlds} in which \emph{artificial people}  behave
similarly to real humans. The rules of these worlds are not obtained
from the ``basic laws'' of the real one, since no computer can
at present simulate the behavior of all the elements of even a small
portion of matter. These rules are designed so to behave similarly
to the system under investigation,
and to be easily implemented in digital terms. There are
two main motivations (or hopes): to be able to
understand real complex
dynamics by studying simplified models, and to be so lucky to discove
r that a finely-tuned model is able to reproduce (or forecast) the
behavior of its real counterpart.

This is so promising that many scientists are 
performing experiments on these artificial world, in order to extract
their principal characteristics, to be subsequently analyzed
possibly
using paper and pencil! 

In the following we shall try to elucidate some aspects of the
interplay between computer and physics. In \secref{DynStat},
we shall
illustrate possible logic pathways (in classical mechanics) leading
from Newton's equations to research fields that use computers as
investigative tool. In \secref{ArtificialWorlds}, we shall try to
present succinctly
some example of ``artificial worlds'' that are still active research
topics
in theoretical physics.

\Section[DynStat]{From Trajectories to Statistics and Back}

Let us assume that a working model of the reality can be built using a
set of dynamical equations, for instance those of classical
mechanics. We shall consider the model of a system formed by
many particles, like a solid or a fluid. The state of the
resulting system can be represented as a point in a high-dimensional
space, since it is given by all coordinates and velocities of all
particles. The evolution of the system is a trajectory in such space.
Clearly, the visualization and the investigation of such a problem is
challenging, even using powerful computers.

Moreover, even if
we are able to compute one or many trajectories (in order to have an
idea of fluctuations), this does not imply that we have
\emph{understood} the problem. Let us consider for instance the
meteorology: one is interested in the probability of rain, or in the
expected wind velocity, not in forecasting the trajectories of all
molecules of air. Similarly in psychology, one is interested in the
expected behavior of an individual, not in computing the activity of
all neurons in his brain.

Since physics is the oldest discipline that has been ``quantified''
into equations, it may be illuminating to  follow some of the paths
followed by  researchers to ``reduce'' the complexity of a
high-dimensional problem to something more manageable, or at least
simpler to be simulated on a computer.

In particular, we shall see that many approaches consist in
``projecting'' the original space onto a limited number of dimensions,
corresponding to the observables that vary in a slow and smooth way,
and assuming that the rest of the dynamics is approximated by
``noise''\footnote{The noise can be so small, compared to the
macroscopic observables that it can be neglected. In such cases, one
has a deterministic, low-dimensional dynamical system, like for
instance the usual models for rigid bodies, planets, etc.}.
Since the resulting system is stochastic, one is interested in
computing the average values of observables, over the probability
distribution of the projected system.

\begin{sidewaysfigure}
\begin{center}
\includegraphics{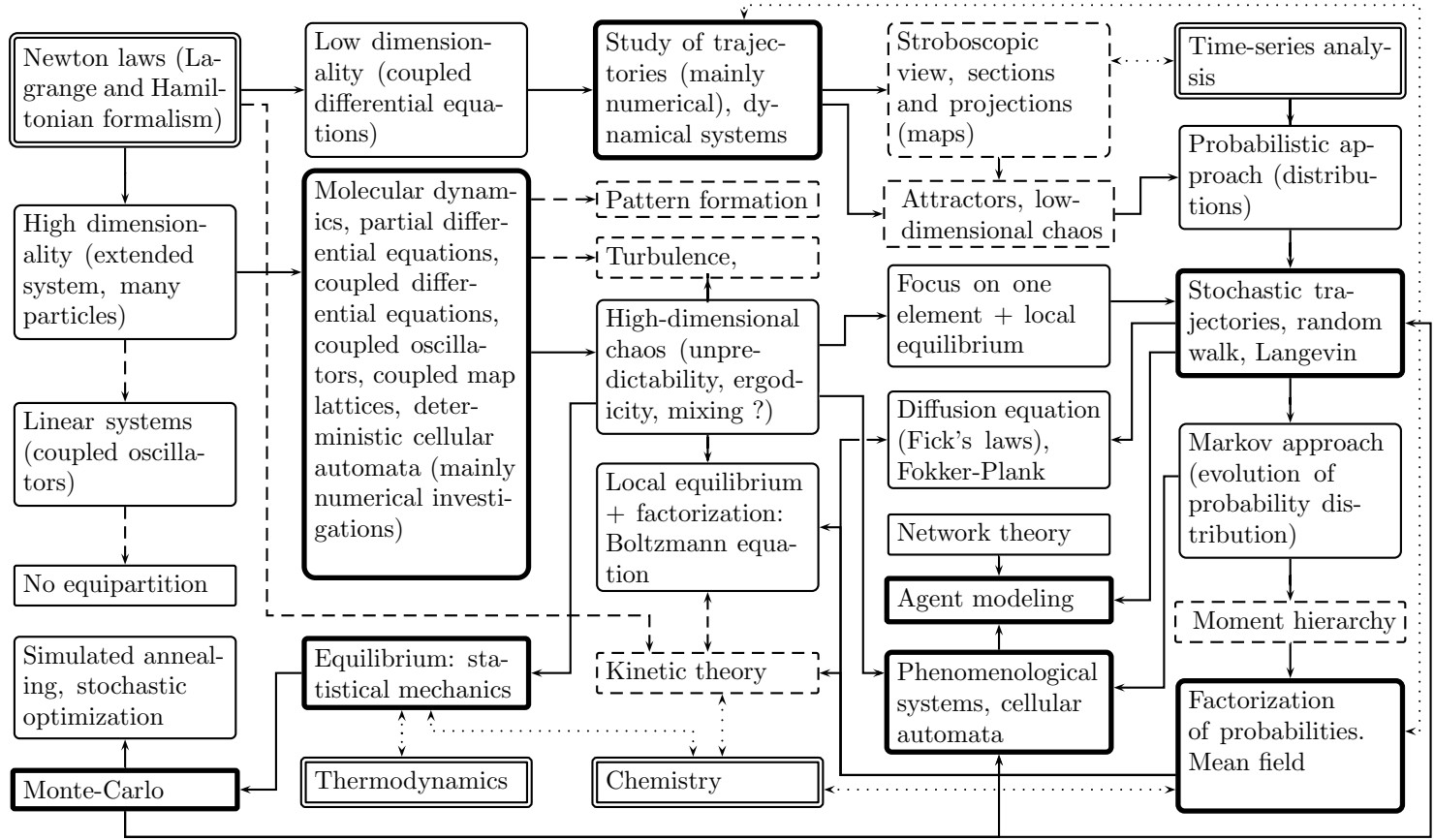}
\end{center}
\caption{\label{scheme} Graphical illustration of the logic path
followed in this introduction. Boxes with double frame are ``starting
points'', dashed boxes are topic that are not covered by discussion,
boxes with darker frames mark topics that are investigated more in
details.}
\end{sidewaysfigure}

However, the computation of the
probability distribution may be hard, and so one seeks to find a way
of producing ``artificial'' trajectories, in the projected space,
designed in such a way that their probability distribution is the
desired one. So doing, the problem reduces to the computation of the
time-averaged values of ``slow'' observables.
For the rest of this section, please make reference to
Figure~\ref{scheme}

\Paragraph{Newton laws}
The success of Newton in describing the motion of one body,
subjected to a static field force (say: gravitational motion of one
planet, oscillation of a body attached to a spring, motion of the
pendulum, etc.) clearly proved the validity of his approach, and also
the validity of using simple models for dealing with natural
phenomena. Indeed, the representation of a body as a point mass,
the idea of massless springs and strings, the concept of  force fields
are all mathematical idealizations of reality. 

The natural generalization of this approach is carried out in the
XVIII century by Lagrange, Hamilton and many others.
It brings to the mathematisation of mechanics and the derivation of
\emph{rational mechanics}. The resulting ``standard'' (or historical)
way of modeling
physical systems is that of using differential equations, \ie, a
continuous description of time, space and other dynamical quantities.

From an abstract point of view, one is faced with two different
options: either  concentrate on systems described by a few equations
(low-dimensional systems), or try to describe systems formed by many
components. 

\Paragraph{Low dimensionality}

Historically, the most important problem of Newton's times was that of
three bodies interacting via gravitational attraction (the Sun, the
Earth and the Moon). By approximating planets with point masses, one
gets a small number of coupled differential equations. This reduction
of dimensionality is an example of a ``scale separation'': the
variables that describe the motion of the planets vary slowly and
smoothly in time. Other variables, for instance those that describe
the oscillations of a molecule on the surface of a planet, can be
approximated by a noise term so small that can be safely neglected.
This approximation can also be seen as a ``mean field'' approach, for
which one assumes that variables behave not too differently from their
average. Using these techniques, one can develop models of many
systems that result in the same mathematical scheme: a few coupled
equations. The resulting equations may clearly have a structure quite
different from that resulting from Newtonian dynamics (technically,
Hamiltonian systems).

However, the reduction of the number of variables does not guarantee
the simplicity of the resulting model. The problem of  three
gravitational bodies cannot be split into smaller pieces, and the
computation of an accurate trajectory requires a computer. In general,
a \emph{nonlinear} system in a space with three or more dimensions is
chaotic. This implies that there it may ``react'' to a small
perturbation of parameters or initial conditions with large variations
of its trajectory. This \emph{sensibility} to variation implies
the impossibility of predicting its behavior for long times, unless
one is content with a probabilistic description.

\Paragraph{High dimensionality}

In many cases, the ``projection'' operation results in a system still
composed by many parts. For instance, models of nonequilibrium fluids
neglect to consider the movement of the individual molecules, but
still one has to deal with the values of the pressure, density and
velocity in all points. In these cases one is left with a
high-dimensional problem. Assuming that the ``noise'' of the projected
dimensions can be neglected, one can either write down a large number
of coupled equation (\eg, in modeling the vibration of a crystal), or
use a continuous approach and describe the system using partial
differential equations (\eg, the model of a fluid).

\Paragraph{Linear systems}
In general, high and low-dimensional approaches can be systematically
developed (with paper and pencil) only in the linear approximation.
Let us illustrate
this point for the case of coupled differential equation: if the
system
is linear one can write the equations using matrices and vectors. One
can in principle find a (linear) transformation of variables that make
the system diagonal, \ie, that reduces the problem to a set of
\emph{uncoupled} equations.  At this point, one is left
with (many) one-dimensional independent problems.
Clearly, there are mathematical difficulties, but the path is clear. 
A similar approach (for instance using Fourier transforms) can be used
also for dealing with partial differential equations. 

The variables that result from such operations are called normal
modes, because they behave independently one from the other (\ie, they
correspond to orthogonal or normal directions in the state space). For
instance, the linear model of a vibrating string (with fixed ends)
predicts that any pattern can be described as a superposition  of
``modes'', which are the standing oscillations with zero, one, two,
\dots nodes (the harmonics).

However, linear systems behave in a somewhat strange way, from the
point of view of thermal physics. Let us
consider for instance the system composed by two \emph{uncoupled}
oscillators. It is
clear that if we excite one oscillator with any amount of energy,
it will remain confined to that subsystem. With normal modes, the
effect is the same: any amount of energy communicated to a normal mode
remains confined to that mode, if the system is completely linear.
In other words, the system never forgets its initial condition.

On the contrary, the long-time behavior of  ``normal'' systems does
not depend strongly on the initial conditions. One example is
the \emph{timbre}, or ``sound color'' of an object. It is given by the
simultaneous oscillations on many frequencies, but in general an
object emits its ``characteristic'' sound regardless of how exactly
is perturbed. This would not be true for linear systems.

Since the \emph{distribution} of energy to all available ``modes'' is
one of the assumptions of equilibrium statistical mechanics, which
allows us to ``understand'' the usual behavior of matter, we arrived
at an unpleasant situation: linear systems, which are so ``easy'' to
be studied, cannot be used to ground statistical physics on
mechanics. 

\Paragraph{Molecular dynamics}
Nowadays, we have computers at our disposal, and therefore we can
simulate systems composed by many parts with complex interactions.
One is generally interested in computing \emph{macroscopic}
quantities. These are defined as \emph{averages} of some
function of the microscopic variables (positions,  velocities,
accelerations, etc.) of the
system. A measurement on a system implies an average,
over a finite interval of time and over a large
number of elementary components (say: atoms, molecules, etc.) of some
quantity that depends on the microscopic state of that portion of the
body. 

\Paragraph{Chaos and probabilities}
It was realized by Poincaré (with paper and pencil) and
Lorenz (with one of the very first computers) that also very
few (three) coupled differential equations with nonlinear interactions
may give origin to complex (chaotic) behavior. In a chaotic system, a
small uncertainty amplifies exponentially in time, making forecasting
difficult. However, chaos may
also be simple: the equations describing the trajectory of dice are
almost surely chaotic, but in this case the chaos is so strong that
the tiniest perturbation or uncertainty in the
initial conditions will cause in a very small amount of time a
complete variation of the trajectory. Our experience says that the
process is well approximated by a probabilistic description.
Therefore, chaos is one possible way of introducing probability in
dynamics.

Chaotic behavior may be obtained in simpler models,
called maps, that evolve in discrete time steps. As May
showed~\cite{May:LogisticMap}, a
simple map with a quadratic non-linearity (logistic map) may be
chaotic. One can also model a system using coupled maps instead of a
system of coupled
differential equations~\cite{Kaneko:CML}. And indeed, when a
continuous system is
simulated on a computer, it is always represented as an array of
coupled maps.

\Paragraph{Discretization}
There is a progression of discretization from partial differential
equations, coupled differential equations, coupled map lattices: from
systems that are continuous in space, time and in the dynamical
variables to
systems that are discrete in time and space, and continuous only in
the dynamical variables. The further logic step is that of studying
completely discrete systems, called \emph{cellular automata}.

Cellular automata show a wide variety of different phenomenologies.
They can be considered mathematical tools, or used to model reality.
In many cases, the resulting phenomenological models follows
probabilistic rules, but
it is also possible to use cellular automata as ``building blocks''.
For instance, is possible to simulate the behavior of a hydrodynamical
system
by means of a completely discrete model, called cellular automata
lattice gas (see \secref{CA}).

\Paragraph{Statistics}
The investigation of chaotic extended systems proceed generally using
a statistical approach. The idea is the following: any system
contains a certain degree of non-linearity, that couples otherwise
independent normal modes. Therefore, (one hopes that) the initial
condition is not too important for the asymptotic regime. If moreover
one assumes that the motion is so chaotic that any trajectory spans
the available space in a ``characteristic'' way (again, not depending
on the initial conditions), we can use statistics to derive the
``characteristic'' probability distribution: the probability of
finding the system in a given portion of the available space is
proportional to the the time that the
system spends in that region. See also the
paragraph on equilibrium. 

\Paragraph{Random walks}
Another approach is that of focusing on a small part of a system, for
instance a single particle. The rest of the system is approximated by
``noise''. This method was applied, for instance, by Einstein
in the development of the simplest theory of Brownian
motion, the
random walk~\cite{EinsteinRandomWalk}. In random walks, each steps of
the target
particle is independent on previous steps, due to collisions with the
rest of particles. Collisions, moreover, are supposed to be
uncorrelated. 
A more sophisticated approximation consists in keeping some aspects of
motion, for instance the influence of inertia or of external forces,
still approximating the rest of the world by noise (which may contain
a certain degree of correlation). This is known as the
Langevin approach, which includes the random walk as the
simplest case. Langevin equations are stochastic differential
equations.  

The essence of this method relies in the assumption that the behavior
of the various parts of the systems is uncorrelated. This assumption
is vital also for other types of approximations, that will be
illustrated in the following. Notice that in the statistical mechanics
approach, this assumption is not required. 

In the Langevin formulation, by averaging over many ``independent''
realizations of the process (which in general is not the same of
averaging over many particles that ``simultaneously'' move, due for
instance to excluded volumes) one obtains the evolution equation of
the probability of finding a particle in a given portion of space.
This is the Kolmogorov integro-differential equation, that in many
case can be simplified, giving a differential (Fokker-Plank) equation.
The
diffusion equation is just  the simplest
case~\cite{VanKampen,Gardiner}.

It is worth noticing that a similar formulation may be developed for
quantum systems: the Feynman path-integral approach is
essentially a Langevin formulation, and the Schroedinger
equation is the corresponding Fokker-Plank equation. 

Random walks and stochastic differential equations find many
application in economics, mainly in market simulations. In this cases,
one is not interested in the average behavior of the market, but
rather in computing non-linear quantities over trajectories
(time-series of good values, fluctuations, etc.).

\Paragraph{Time-series data analyses}
In practice, a model is never derived ``ab initio'', by projecting the
dynamics of all the microscopic components onto a limited number of
dimensions, but is constructed heuristically from observations of the
behavior of a real system. 

It is therefore crucial to investigate how observations are made, \ie,
the analysis of a series of a time measurements. In particular, a good
exercise is that of simulating a dynamical or stochastic system,
analyze the resulting time-series data of a given observable, and see
if one is able to reconstruct from it the relationships or the
equations ruling the time evolution.

Let us consider the experimental study of a chaotic,
low-dimensional system. The measurements on this system give a time
series of values, that we assume discrete (which is actually the case
considering experimental errors). Therefore, the output of our
experiment is a series of symbols or numbers, a time-series.
Let us assume that the system is stationary, \ie, that the sequence
is statistically homogeneous in time.  If the system is not extremely
chaotic, symbols in the sequence are correlated, and one can derive
the probability of observing single symbols, couples of symbols,
triples
of symbols and so on. There is a hierarchy in these probabilities,
since the knowledge of the distribution of triples allows the
computation of the distribution of couples, and so on. 

It can be shown that the knowledge of the probability distribution of
the infinite sequence is equivalent to the complete knowledge of the
dynamics. However, this would correspond to performing an infinite
number of experiments, for all possible initial conditions.  

The usual investigation scheme assumes that
correlations vanishes beyond a
certain distances, which is equivalent to assume that the
probability of
 observing sequences longer than that distance factorize. 
Therefore, one tries to model the evolution of the
system by a probabilistic dynamics of symbols~\secref{DP}.
Time-series data analysis can therefore be considered as the main
experimental motivation in developing probabilistic discrete models. 
This can be done heuristically comparing results with observations
\emph{a posteriori}, or trying to extract the rules directly from
data, like in the Markov approach.

\Paragraph{Markov approximation}

The Markov approach, either continuous or discrete, also
assumes that the memory of the system vanishes after a certain time
interval, \ie, that the correlations in time series decay
exponentially. In discrete terms, one tries to describe the process
under study as an automata, with given transition
probabilities. The main problem is:
given a sequence of symbols, which is the simplest automata
(hidden Markov chains~\cite{HiddenMarkovModel}) that can generate
that sequence with maximum ``predictability'', \ie, with transition
probabilities that
are nearest to zero or one? Again, it is possible to derive  a
completely
deterministic automata, but in general it has a number of nodes
equivalent to the length of the time-series, so it is not
generalizable and has no predictability. On the contrary, an automata
with a very small number of nodes will have typically intermediate
transition probabilities, so predictability is again low (essentially
equivalent to random extraction). Therefore, the good model is the
result of an
optimization problem, that can be studied using, for instance,
Monte-Carlo techniques. 

\Paragraph{Mean-field}
Finally, from the probabilities one can compute averages of
observables, fluctuations and other quantities called ``moments'' of
the distribution. Actually, the knowledge of all moments is equivalent
to
the knowledge of the whole distribution. Therefore, another approach
is that of relating moments at different times or different locations,
truncating the recurrences at a certain level. The roughest
approximation is that of truncating the relations at the level of
averages, which is the mean field approach. It appears so
natural that is is often used without realizing the implications of
the approximations. For instances, chemical equations are essentially
mean-field approximations of a complex phenomena. 

\Paragraph{Boltzmann equation}
Another similar approach is that of dividing an extended system into
zones, and assume that the behavior of the system in each zone is well
described by a probability distribution. By disregarding correlations
with other zones, one obtains the Boltzmann equation, with
which many transport phenomena may be studied well beyond elementary
kinetic theory. The Boltzmann equation can also be obtained from the
truncation of a hierarchy of equations (BBGKY hierarchy) relating
multi-particle probability distributions. Therefore, the Boltzmann
equations is similar in spirit to a mean-field analysis.

\Paragraph{Equilibrium}
One of the biggest success of the stochastic approach is of course
\emph{equilibrium statistical mechanics}. The main ingredient of this
approach is that of minimum information, which, in other words,
corresponds to the assumption: \emph{what
is not known is not harmful}. By supposing that in equilibrium the
probability distribution of the systems maximizes the information
entropy
(corresponding to a minimum of information on the system) one is
capable of deriving the probability distribution itself and therefore
the expected values of observables (ensemble averages, see
\secref{Ising}). In this way,
using an explicit model, one is capable to compute the value of
the parameters that appear in thermodynamics. If  it were
possible to show that the maximum entropy state is actually the state
originated by the dynamics of a mechanical (or quantum) system, one
could ground thermodynamics on mechanics. This is a long-investigated
subject, dating back to Boltzmann, which is however not yet clarified.
The main drawback in the derivations is about ergodicity. Roughly
speaking, a system is called \emph{ergodic} if the infinite-time
average of an observable over a trajectory coincides with its average
over a snapshot of
infinitely many replicas. For a system with fixed energy
and no other conserved quantities, a sufficient condition is that a
generic trajectory passes ``near'' all points of the  accessible
phase-space. However, most systems whose behavior is
``experimentally'' well approximated by statistical mechanics are not
ergodic. Moreover, another ingredient, the capability of
\emph{forgetting}
quickly the information about initial conditions appears to be
required, otherwise trajectories are strongly correlated and averages
over diferent trajectories cannot be ``mixed'' together. This
capability is
strongly connected to the chaoticity or \emph{unpredictability} of
extended systems, but unfortunately these ingredients makes
analytic approaches quite hard. 

An alternative approach, due to Jaynes~\cite{Jaynes:MaximumEntropy},
is much more pragmatic. In essence, it says: let design a model with
the ingredients that one thinks are important, and assume that all
what is not in the model does not affect its statistical properties.
Compute the distribution that maximizes the entropy with
the given constraints. Then, compare the results (averages of
observables) with experiments (possibly, numerical ones). If they
agree, one has captured the essence of the problem, otherwise one have
to
include some other ingredient and repeat the procedure. Clearly,
this approach is much more general than the ``dynamical'' one, not
considering trajectory or making assumptions about the energy, which
is simply seen as a constraint. But physicists would be much more
satisfied by a ``microscopic'' derivation of statistical physics.

In spite of this lack of strong basis, the statistical mechanics
approach is  quite powerful, especially for systems that can be
reduced to the case of \emph{almost} independent elements. In this
situation,
the system (the \emph{partition function}) factorizes, and many
computations may be performed by hand. Notice however that this
behavior is in strong contrast to that of truly linear systems: the
``almost'' attribute indicates that actually the elements interact,
and
therefore share the same ``temperature''. 

\Paragraph{Monte-Carlo}
The Monte-Carlo technique was invented for computing, with the aid of
a computer, thermal averages of observables of physical systems at
equilibrium. Since then, this term is often used to denote the
technique of computing the average values of observables of a
stochastic system by computing the time-average values over
``artificial'' trajectories.

In equilibrium statistical physics, one is faced by the problem of
computing averages of observables over the probability distribution of
the system, and since the phase space is very high-dimensional, this
is in general not an easy task: one cannot simply draw \emph{random}
configurations, because in general they are so different from those
``typical'' of the given value of the temperature, that their
statistical weight is marginal. And one does not want to revert to the
original, still-more-highly-dimensional dynamical system, which
typically requires powerful computers just to be followed for tiny
time intervals.

First of all, one can divide (\emph{separate}) the model into almost
independent subsystems, that, due to the small residual interactions
(the ``almost'' independency), are at the same temperature. In the
typical example of a gas, the velocity components appear into the
formula of energy as additive terms, \ie, they do not interact with
themselves or with other variables. Therefore, they can be studied
separately giving the \emph{Maxwell distribution} of velocities. The
positions of molecules, however, are linked by the potential energy
(except in the case of an ideal gas), and so the hard part of the
computation is that of generating configurations. Secondly,
statistical mechanics guarantees that the asymptotic probability
distribution does not depend on the details of dynamics. Therefore,
one is free to look for the fastest dynamics still compatible with
constraints. The Monte-Carlo computation is just a set of recipes for
generating such trajectories. In many problems, this approach allows
to reduce the (computational) complexity of the problem of several
orders of magnitude, allowing to generate ``artificial'' trajectories
that span statistically significant configuration with small
computational effort. In parallel with the generation of the
trajectory, one can compute the value of several observables, and
perform statistical analysis on them, in particular the computation of
time averages and fluctuations. 

By extension, the same terms ``Monte-Carlo'' is used for the technique
of generating sequences of states (trajectories) given the transition
probabilities, and computing averages of observables on trajectories,
instead of on the probability distribution.

One of the most interesting applications of Monte-Carlo simulations
concerns stochastic optimization via \emph{simulated annealing}. The
idea is that of making an analogy between the status of a system (and
its energy) and the coding of a particular procedure (with
corresponding cost function). The goal is that of finding the best
solution, \ie,  the global minimum of the energy. ``Easy'' system
have
a smooth energy landscape, funnel shaped, so that usual techniques
like gradient descent are successful. However, when the energy
landscape is corrugated, there are many local minima where local
algorithms tend to be trapped. Methods from statistical mechanics
(Monte-Carlo), on the contrary, are targeted to generate trajectories
with short correlations, in order to allow the system to quickly
explore all available space. By lowering the ``temperature'', the
probability distribution of system obeying statistical mechanics
concentrate around minima of energy, and a good Monte-Carlo trajectory
should do the same. Therefore, a sufficiently slow annealing
should
furnish the desired global minimum. 

\Paragraph{Critical phenomena}
One of the most investigated topics of statistical mechanics concerns
phase transitions. This is a fascinating subject: in the
vicinity of a continuous phase transitions correlation lengths
diverge, and the system behave collectively, in a way which is largely
independent of the details of the model. This universal
behavior allows the use of extremely simplified models, that therefore
can be massively simulated. 

The \emph{philosophy} of statistical mechanics may be
\emph{exported} to nonequilibrium systems: systems with absorbing
states
(that correspond to infinitely negative energy), driven systems (live
living ones), strongly frustrated systems (that actually never reach
equilibrium), etc. In general, one defines these systems in terms of
transition probabilities, not in term of energy. Therefore, one cannot
invoke a maximum energy principles, and the results are less
general. The study is generally performed by a mix of analytic
approaches, from Markov chains to stochastic differential equations to
\emph{ad-hoc} techniques (conformal invariance, damage spreading,
replicas, synchrony etc.).

However, many system exhibit behavior reminiscent of equilibrium
systems, and the same language can be used: phase transitions,
correlations, susceptibilities,...These characteristics,
common to many different models, are sometimes referred as emergent
features.

One of the most famous problems in this field is percolation:
the formations of giant clusters in systems described by a local
stochastic aggregation dynamics. This ``basic'' model has been used to
describe an incredibly large range of
phenomena~\cite{Stauffer:Percolation}.

Equilibrium and nonequilibrium phase transitions occur for a
well-defined value of a control parameter. However, in nature one
often observes phenomena whose characteristic resemble that of a
system near a phase transition, a critical dynamics, without
any \emph{fine-tuned} parameter. For such system the term
self-organized criticality has been coined~\cite{BTW:SOC}, and they
are
subject of active researches. 

\Paragraph{Networks}
A recent ``extension'' of statistical physics is the theory of
networks.
Networks in physics are often regular, like the 
contacts in a crystal, or only slightly randomized. Characteristics of
these networks are the fixed (or slightly dispersed around the mean)
number of connections per node, the high probability of having
connected neighbors (number of ``triangles''), the large time needed
to
cross the network. The opposite of a regular network is a random
graph, which, for the same number of connections, exhibit low number
of triangles, short crossing time. The statistical and dynamical
properties of systems whose connection are regular or random are
generally quite different. 

Watts and Strogatz~\cite{WattsStrogatz} argued that  \emph{social
networks}  are never
completely regular. They showed that the simple
random rewiring
of a small number of links in a regular network may induce the
small world effect: local properties, like the number of
triangles, are not affected, but large-distance ones, like the
crossing time, quickly became similar to that of random graphs. Also
the statistical and dynamical properties of models defined over a
rewired networks are generally similar to those correlated to random
graphs. 

After this finding, many social networks were studied, and they
revealed a yet different structure: instead of having a well-defined
connectivity, many of them present a few highly-connected ``hubs'',
and
a lot of poorly-connected ``leafs''. The distribution of connectivity
of
often distributed as a power-law (or similar~\cite{Newman:PowerLaws}), without a well-defined mean
connectivity (scale-free networks~\cite{BarabasiAlbert:ScaleFree}).
Many of phenomenological
models are presently re-examined in order to investigate their
behavior if defined over such networks. Moreover, scale-free networks
cannot be ``laid down'', they need to be ``grown'' following a
procedure
(similar in this to fractals). It is natural therefore to
include such a procedure in the model, design therefore models that
not only evolve ``over'' the networks, but also evolve ``the''
network~\cite{ComplexDynamics}.

\Paragraph{Agents}
Many of the described tools are used is the so-called
agent-based models. The idea is that of exploiting the
powerful capabilities of present computers to simulate directly a
large number of \emph{agents} that interact among them. Traditional
investigations of \emph{complex systems}, like crowds, flocks,
traffic,
urban models, and so on, have been performed using \emph{homogeneous}
representation: partial differential equations (\ie, mean-field),
Markov equations, cellular automata, etc. In such an approach, it is
supposed that each agent type is present in many identical copies, and
therefore they are simulated as ``macrovariables'' (cellular
automata),
or aggregated like independent random walkers in the diffusion
equation. But live elements (cells, organisms) do not behave in such a
way: they are often individually unique, carry information about their
own past history, and so on. With computers, we are now in the
position of simulating large assemblies of individuals, possibly
geographically located (say, humans in an urban simulation).

One of the advantages of such approach is that of offering the
possibility of measuring quantities that are inaccessible to
experimentalists, and also to experiment with different scenarios. The
disadvantages are the proliferation of parameters, that are often
beyond experimental confirmation.

\Section[ArtificialWorlds]{Artificial worlds}

A somewhat alternative approach to that of ``traditional''
computational physics is that of studying an artificial model, build
with little or no direct connection with reality, trying
to include only those aspect that are considered relevant. The goal
is to be able to find the \emph{simplest} system still able to exhibit
the relevant features of the phenomena under investigation. For
instance, this is the spirit with which the Ising model was build.

\Subsection[Ising]{Ising}

The Ising (better: Lenz-Ising) model is probably one of the most known
models in statistical physics. Its history~\cite{HistoryIsingModel}
is particularly illuminating in this context, even if it happened well
before the advent of computers in physics.
 
Let us first illustrate schematically the model. It is
defined on a lattice, that can be in one, two or more dimensions, or
even on a disordered graph. We shall locate a cell with
an index ${\vec{i}}$, corresponding to the set of spatial
coordinates for a regular lattice or a label for a graph. The
dynamical variable $x_{\vec{i}}$ for each cell
is just a binary digit, traditionally named ``spin'' that takes the
values  $\pm1$. 
We shall indicate the whole configuration as $\vec{x}$. 
Therefore, a lattice with $N$ cells has $2^N$ distinct
configurations. Each configuration $\vec{x}$ has an associated
energy 
\[
  E(\vec{x}) = -\sum_{\vec{i}} (H+h_{\vec{i}}) x_{\vec{i}},
\]
where $H$ represents the external magnetic field and $h_{\vec{i}}$ is
a local magnetic field,
generated by neighboring spins,
\[
 h_{\vec{i}} = \sum_{\vec{j}} J_{{\vec{i}}{\vec{j}}} x_{\vec{i}}.
\]
The coupling $J_{{\vec{i}}{\vec{j}}}$ for the original Lenz-Ising
model is
\eq[J]{
  J_{{\vec{i}}{\vec{j}}} = \begin{cases} J &\text{if ${\vec{i}}$ and
${\vec{j}}$ are first        neighbors,}\\
      0 & \text{otherwise.}\end{cases}
}
The spins belonging to the neighborhood of cell ${\vec{i}}$
($J_{{\vec{i}}{\vec{j}}}>0$) are indicated as $\vec{X}_{\vec{i}}$.

The
maximum-entropy principle~\cite{Jaynes:MaximumEntropy} gives the
probability distribution
\[
 P(\vec{x}) = \frac{1}{Z}\exp\left(-\frac{E(\vec{x})}{T} \right)
\]
from which averages can be computed. The parameter $T$ is the
temperature, and $Z$, the ``partition function''
is the normalization factor of the distribution.

The quantity $E(\vec{x})$ can be though as a
``landscape'', with low-energy configurations corresponding to valley
and high-energy ones to peaks. The distribution $P(\vec{x})$ can be
interpreted as the density of a gas, each ``particle'' corresponding
to a possible realization (a replica) of the system. This \emph{gas}
concentrates in the valleys for low temperatures, and diffuses if
the temperature is increased. The temperature is related to the
average level of the gas.

In the absence of the local field ($J=0$), the energy is
minimized if each $x_{\vec{i}}$ is \emph{aligned} (same sign) with
$H$.
This
ordering is counteracted by thermal noise. In this case it is
quite easy to obtain  the average magnetization per spin (order
parameter)
\[
 \langle
x \rangle= \tanh\left(\frac{H}{T}\right),
\]
which is a plausible behavior for a paramagnet. A ferromagnet however
present hysteresis, \ie, it may maintain for long times
(metastability) a pre-existing magnetization opposed to the external
magnetic
field. 

With coupling turned
on ($J>0$), it may happen that the local field is strong enough to
``resist'' $H$, \ie, a compact patch of spins oriented against $H$ may
be stable, even if the energy could be lowered by flipping all them,
because the flip of a single spin would rise the energy (actually,
this flip may happen, but is statistically re-absorbed in short
times). The fate of the patch is governed by boundaries. A spin on a
boundary of a patch feels a weaker local field, since some of its
neighbors are oriented in the opposite. Straight boundaries in two or
more dimensions separate spins that ``know'' the phase they belong
to, since most of their neighbors are in that phase, the spins on the
edges may flip more freely. Stripes that span the whole
lattice are rather stable objects, and may \emph{resist} an opposite
external field since
spins that occasionally flip are surrounded by spins belonging to the
opposite phase, and therefore feel a strong local field that pushes
them towards the phase opposed to the external field. 
 
In one dimensions with finite-range coupling, a single spin flip is
able to create a ``stripe'' (perpendicularly to the the lattice
dimension), and therefore can destabilize the ordered phase. This is
the main reason for the absence of phase transitions in 
one dimension, unless the coupling extends on very large
distances or some coupling is infinite (see the part on directed
percolation, \secref{DP}).

This model was proposed  in the early 1920s by Lenz to
Ising for his PhD dissertation as a simple model of a ferromagnet.
Ising studied it in one dimension, found that it shows no phase
transition and concluded (erroneously) that the same happened in
higher dimensions. Most of contemporaries rejected the model since it
was not based on Heisenberg's quantum mechanical model of
ferromagnetic interactions. It was only in the forties that it
started gaining popularity as a model of cooperative phenomena, a
prototype of order-disorder transitions. Finally, in 1940,
Onsager~\cite{Onsager:Ising2D} provided the exact solution of the
two-dimensional Lenz-Ising model in zero external field. It was the
first (and for many year the only) model exhibiting a non-trivial
second-order transition whose behavior could be exactly computed. 

Second-order transition have interested physicists for almost all the
past century. In the vicinity of such transitions, the elements (say,
 spins) of the system are correlated up to very large distances. For
instance, in the Lenz-Ising model (with coupling and more than one
dimension), the high-temperature phase  is disordered,
and the low-temperature  phase is almost completely
ordered. In both these phases the connected two-points correlation
function
\eq{
  G_c(r) = \langle x_{\vec{i}} x_{{\vec{i}}+\vec{r}}\rangle -
\langle
x_{\vec{i}}\rangle^2 
}
decreases exponentially, $G_c(r)\simeq
\exp(-r/\xi)$, with $r=|{\vec{r}}|$. The length $\xi$ is a measure of
the typical size of
patch of spins pointing in the same direction. 

Near the critical temperature $T_c$, the
correlation length $\xi$ diverges like
$\xi(T-T_c)\simeq (T-T_c)^{-\nu}$ ($\nu=1$ for $d=2$ and $\nu\simeq
 0.627$ for $d=3$, where $d$ is the dimensionality of the system). In
such case the correlation function is described by a power law
\eq{
  G_c(r) \simeq \frac{1}{r^{d-2+\eta}},
}
with $\eta=1/4$ for $d=1$ and $\eta\simeq 0.024$ for $d=3$.
This phase transition is an
example of a \emph{critical
phenomenon}~\cite{TheoryOfCriticalPhenomena}, $\nu$ and $\eta$ are
examples of \emph{critical exponents}.

The divergence of the correlation length indicates that
there is no characteristic scale ($\xi$) and therefore
fluctuations of all sizes appear.
In this case,
the details of the interactions are not so important, so that many
different models behave in the same way, for what concerns for
instance the critical exponents. Therefore, models can be grouped into
\emph{universality classes}, whose details are essentially given by
``robust'' characteristics like the dimensionality of space and of
the order parameter, the symmetries, etc.

The power-law behavior of the correlation function also indicates that
if we perform a \emph{rescaling} of the system, it would appear the
same or, conversely,  that one is unable to estimate the
\emph{distance}
of a pattern by comparing the ``typical size'' of particulars.
This scale invariance if typical of many natural phenomena, from
clouds (whose height and size are hard to be estimated), trees and
other plant elements, lungs, brain, etc.

Many examples of power laws and collective behavior can be found in
natural sciences~\cite{Sornette:CriticalPhenomena}. Differently from
what happens in the Lenz-Ising model, in these cases there is no
parameter (like the temperature) that has to be fine-tuned,
so one speaks of \emph{self-organized criticality}~\cite{BTW:SOC}. 

Since the Lenz-Ising model is so simple, exhibits a critical phase
and can be exactly solved (in some case), it has become the
playground for a variety of modifications  and applications
to various fields. Clearly, most of modifications do no allow
analytical treatment and have to be investigated
numerically. The Monte-Carlo method allows to add
a temporal dimension to a statistical
model~\cite{Kawasaki:DynamicIsing}, \ie, to transform
stochastic integrals into averages over fictitious trajectories.
Needless to say, the Ising-Lenz model is the standard test for every
Monte-Carlo beginner, and most of techniques for accelerating the
convergence of averages has been developed with this model in
mind~\cite{SwendsenWang}. 

Near a second-order phase transitions, a physical system
exhibits \emph{critical slowing down}, \ie, it reacts to external
perturbations with an extremely slow dynamics, with a convergence
time that increases with the system size. One can extend the
definition of the correlation function including the time dimension:
in the critical phase also the temporal correlation length diverges
(as a power law). This happens also for the
Lenz-Ising model using the the Monte-Carlo dynamics, unless very
special techniques are used~\cite{SwendsenWang}. Therefore, the
dynamical version of the Lenz-Ising model can be used also to
investigate relaxational dynamics and how this is influenced by
the characteristics of the energy landscape. In particular, if the
coupling $J_{\vec{i}\vec{j}}$ changes
sign randomly for
every couple of sites (or the field $H$ has random sign for each
site), the energy landscape becomes extremely rugged. When spins flip
in order to align to the local field, they may invert the field
felt by neighboring ones. This \emph{frustration} effect is believed
to be the basic mechanism of the large viscosity and memory effects of
\emph{glassy}
substances~\cite{MezardParisiVirasoro:SpinGlass,Dotsenko:SpinGlass}. 

The rough energy landscape of glassy systems is also challenging for
optimization methods, like \emph{simulated
annealing}~\cite{Kirkpatrick:SimulatedAnnealing} and its ``improved''
cousin, \emph{simulated
tempering}~\cite{MarinariParisi:SimulatedTempering}. Again,
the Lenz-Ising model is the natural playground for these algorithm. 

The dynamical Lenz-Ising model can be formulated such that each spin
is updated in parallel~\cite{ParallelIsing} (with the precaution of
dividing cells into sublattices, in order to keep the neighborhood of
each cell fixed during updates). In this way, it is
essentially a probabilistic cellular automata, as illustrated
in\secref{DP}.

\Subsection[CA]{Cellular Automata}

In the same period in which traditional computation was
developed, in the early fifties, John Von Neumann  was interested in
the logic basis of life and in particular in self-reproduction, and
since the analysis of a self-reproduction automata following the
rules of real physics was too difficult, he designed a playground (a
cellular automaton) with just enough ``physical rules'' in order to
make its analysis possible. It was however just a theoretical
exercise, the automaton was so huge that up to now it has not yet
completely implemented~\cite{UniversalConstructorUrl}.

The idea of cellular automata is quite simple: take a lattice (or a
 graph) and put on
each cell an automaton (all
automata are equal). Each automaton exhibit its ``state'' (which is
one out of a small number) and is programmed so to react (change
state) according to the state of neighbors, and its present one (the
\emph{evolution rule}). All
automata update their state synchronously.

Cellular automata 
share many similarities with the parallel version of the
Lenz-Ising model. Differently from that, their dynamics is not
derived from an energy, but is defined in terms of the transition
rules. These rules may be deterministic or probabilistic. In the first
case (illustrated in this section), cellular automata are
\emph{fully discrete, extended dynamical systems}. Probabilistic
cellular automata are illustrated in \secref{DP}.
The temporal evolution of deterministic cellular automata can be 
computed \emph{exactly} (regardless of any approximation) on a
standard computer. 

Let us illustrate the simplest case, \emph{elementary cellular
automata}, in Wolfram's jargon~\cite{Wolfram:StatisticalMechanicsCA}.
The
lattice is here one dimensional, so to identify an automaton it is
sufficient to give one coordinate, say $i$, with $i=1, \dots, N$. The
state of the automaton on cell $i$ at time $t$  is
represented by a single variable, $x_i(t)$ that can take only two
values, ``dead/live'', or ``inactive/active'' or 0/1. The time is
also discrete, so $t=1,2,\dots$. 

The parallel evolution of each automaton is given by the rule
\[
  x_i(t+1) = f(x_{i-1}(t), x_i(t), x_{i+1}(t)).
\]
Since $x_i=0,1$, there are only eight possible combinations of
the triple $\{x_{i-1}(t), x_i(t), x_{i+1}(t)\}$, from $\{0,0,0\}$ to
$\{1,1,1\}$. For
each of them, $f(x_{i-1}(t), x_i(t), x_{i+1}(t))$ is either zero or
one, so the function $f$ can be simply coded as a vector of eight
bits, each position labeled by a different configuration of inputs.
Therefore, there are only $2^8=256$ different elementary cellular
automata, that have been studied carefully (see for instance
Ref.~\cite{Wolfram:StatisticalMechanicsCA}). 

\begin{figure}
\begin{center}
\includegraphics[width=0.8\columnwidth]{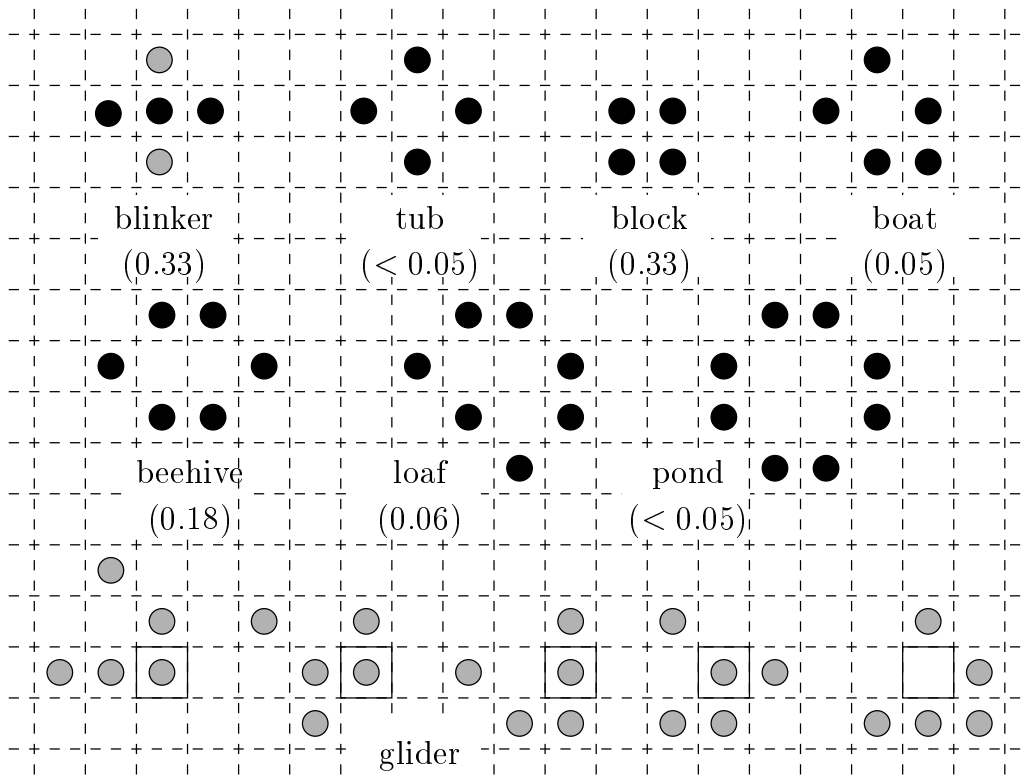}
\end{center}
\caption{\label{Life}Some of the most common ``animals'' in the
game of life, with the probability of encountering them in an
asymptotic  configuration~\protect\cite{Bagnoli:Life}.}
\end{figure}

In spite of their simplicity, elementary cellular automata exhibit a
large variety of behaviors. In the language of dynamical systems they
can be ``visually'' classified~\cite{Wolfram:StatisticalMechanicsCA}
as
fixed points (class-1), limit cycles (class-2) and ``chaotic''
oscillations (class-3). A fourth
class, ``complex'' CA, emerges with
larger neighborhood or in higher dimensions. A classical example is
the
\emph{Game of Life}~\cite{Conway:GameOfLife}. This two-dimensional
cellular automaton is based on a simple rule. A cell may be either
empty (0) or ``alive'' (1). A living cell survives if, among its
8 narest neighbors, there are two or three alive cells, otherwise it
dies and disappears. Generation is implemented through a rule for
empty cells: they may become ``alive'' if surrounded by exactly
three living cells. In spite of the simplicity of the rule, this
automaton generates complex and long-living patterns, some of them
illustrated in Figure~\ref{Life}.

Complex CA
have large transients, during which interesting structures emerge,
after which they relax to class-1 automata. It has been conjectured
that they are able of computation, \ie, that one can ``design'' an
universal computer using these CA as building blocks, as has been
proved to be possible with the Game of Life. Another hypothesis,
again confirmed by the Game of Life, is that these automata are
``near the edge'' of self-organizing complexity. One can slightly
``randomize'' the Game of Life, allowing sometimes an exception to
the rule. Let us introduce a parameter $p$, that measures this
randomness, with the assumption that $p=0$ is the true ``life''.
Well, it was shown~\cite{LifeSOC} that the resulting model exhibits a
second-order phase transition for a value of $p$ very near zero.

Deterministic cellular automata have been investigated as prototypes
of discrete dynamical systems, in particular for what concerns the
definition of chaos. Visually, one is tempted to use this word also
to denote the irregular behavior of ``class 3'' rules. However, the
usual definition of chaos involves the sensitivity to an
infinitesimally small perturbation: following the time dynamics of two
initially close configurations one can observe an amplification of
their distance. If the initial distance ($\delta_0$) is infinitesimal,
then the distance grows exponentially for some time ($\delta(t)\simeq
\delta_0 \exp(\lambda t)$), after which it tends to saturate (since
the trajectories are generally bounded inside an attractor, or due to
the dimensions of the accessible space. The exponent $\lambda$
depends on the initial configuration, and if this behavior is
observed for different portions of the trajectory, it fluctuates: a
trajectory spends some time in regions of high chaoticity, after
which may pass through ``quiet'' zones.
If one ``renormalizes'' periodically this distance, considering one
system as the ``master'' and the other as a measuring device, one can
accumulate good statistics, and define a Lyapunov exponent $\lambda$,
that gives indications about the chaoticity of the trajectory, 
through a limiting procedure.

The accuracy of computation poses some problems. Since in a
computer numbers are always approximate, one cannot follow ``one''
trajectory. The small approximations accumulates exponentially, and
the computer time series actually \emph{jumps} among neighboring
trajectories. Since the Lyapunov exponent is generally not so
sensible to a change of precision in computation, one can assume that
the chaotic regions are rather compact and uniform, so that in
general one associate a Lyapunov exponent to a system, not to an
individual trajectory. Nevertheless, this definition cannot apply to
completely discrete systems like cellular automata.

However, chaoticity is related to unpredictability. As first observed
by Lorenz, and following the definition of Lyapunov exponent, the
precision of an observation over a chaotic system is related to the
average time for which predictions are possible. Like in weather
forecasts, in order to increase the time span of a prediction one has
to increase the precision of the initial measurement. In extended
system, this also implies to extend the measurements over a larger
area. One can also consider a ``synchronization'' approach. Take two
replicas of a systems and let them evolve starting from different
initial configurations. With a frequency and a strength that depends
on a parameter $q$, one of these replica is ``pushed'' towards the
other one, so to reduce their distance. Suppose that $q=0$ is the
case of no push and $q=1$ is the case of extremal push, for which the
two systems synchronize in a very short time. There should be a
critical value $q_c$ that separates these two behaviors (actually,
the scenario may be more complex, with many
phases~\cite{BagnoliCecconi}). In the vicinity of $q_c$ the distance
between the two replicas is small, and the distance $\delta$ grows in
exponential way. The critical value $q_c$ is such that the
exponential growth exactly compensates the shrinking factor, and is
therefore related to the Lyapunov exponent $\lambda$. 

Finite-size cellular automata always follow periodic
trajectories. Let us consider for instance a Boolean automata, of $N$
cells. The number of possible different states is $2^N$ and due to
determinism, once that a state has been visited twice the automata
has entered a limit cycle (or a fixed point). One may have limit
cycles with large basins of transient configurations (configurations
that do not belong to the cycle). Many scenarios are
possible. The set of different configurations may be divided in many
basins, of small size (small transient) and small period, like in
class 1 and 2 automata. Or one may have large basins, with long
transients that lead to short cycles, like in class-4 automata.
Finally, one may have one or very few large basins, with long cycles
that includes most of configurations belonging to the basin (small
transients). This is the case of class-3 automata. For them, the
typical period of a limit cycle grows exponentially (like the
total number of configurations) with the system size, so that for
moderately large system it is almost impossible to observe a whole
cycle in a finite time. Another common characteristic of class-3
automata is that the configurations quickly decorrelates (in the
sense of the correlation function) along a trajectory. If
one takes into consideration as starting points two configurations
that are the same except for a local difference, one observes that
this difference amplifies and diffuses in class-3 automata, shrinks
or remains limited in class-1 and class-2, and have an erratic
transient behavior in class-4, followed by the fate of class-1 and 2. 
Therefore, if one considers the possibility of not knowing exactly
the initial configuration of an automata, unpredictability
grows with time also for such discrete systems. Actually, also
(partially) continuous systems like coupled maps may exhibit this
kind of
behavior~\cite{BagnoliCecconi,CrutchfieldKaneko:StableChaos,
StableChaos,PolitiLiviCecconi:StableChaos}.
Along this line,
it is possible to define an equivalent of the Lyapunov
exponents for CA~\cite{Bagnoli:LyapunovCA}. The synchronization
procedure can be applied also to cellular automata, and it correlates
well with the Lyapunov exponents~\cite{Bagnoli:SynchroCA}.

An ``industrial'' application of cellular automata is their use for
modeling gases. The \emph{hydrodynamical equations}, like the
Navier-Stokes ones, simply reflect the conservation of mass, momentum
and energy (\ie, rotational, translational and time invariance) for
the microscopic collision rules among particles. Since the modeling
of a gas via molecular dynamics is rather cumbersome, some years ago
it was proposed~\cite{HPP,FHP} to simplify drastically the
microscopic dynamics using particles that may travel only along
certain directions with some discrete velocities and jumping in
discrete time only among nodes of a lattice. Indeed, a cellular
automaton. It has been shown that their
macroscopic dynamics is described by usual hydrodynamics laws (with
certain odd feature due to the underlining lattice and finiteness of
velocities)~\cite{RothmanZaleski:LGCA,IntroductionLGCALBE}.

The hope was that these Lattice Gas Cellular Automata
(LGCA) could be simulated so efficiently in hardware to make possible
the investigation of turbulence, or, in other words, that they could
constitute the Ising model of hydrodynamics. While they are indeed
useful to investigate certain properties of gases (for instance,
chemical reactions~\cite{ReactiveLatticeGas}, or the relationship
between chaoticity and equilibrium~\cite{Bagnoli:LGCA}),
they resulted too noisy and too viscous to be useful for the
investigation of turbulence. Viscosity is
related to the transport of momentum in a direction perpendicular to
the momentum itself. If the collision rule does not ``spread''
quickly the particles, the viscosity is height. In LGCA there are many
limitations to collisions, so that in order to lower viscosity one
has to consider averages over large patches, thus lowering the
efficiency of the method.

However, LGCA inspired a very interesting approximation. Let us
consider a large assembly of replicas of the same system, each one
starting from a different initial configuration, all compatible with
the same macroscopic initial conditions. The macroscopic behavior
after a certain time would be the \emph{average} over the status of
all these replicas. If one assumes a form of
local equilibrium, \ie, applies the mean-field approximation for a
given site, one may try to obtain the dynamics of the \emph{average}
distribution of particles, which in principle is the same of
``exchanging'' particles that happen to stay on the same node among
replicas.
 
It is possible to express the dynamics of the average distribution in
a simple form: it is the Lattice Boltzmann Equation
(LBE)~\cite{IntroductionLGCALBE,Succi:LatticeBoltzmann,%
Chopard:CAAndLB}.
The method retains many properties of LGCA like the possibility of
considering irregular and varying boundaries, and may be simulated in
a very efficient way with parallel
machines~\cite{Succi:LatticeBoltzmann}. Differently
from LGCA, there are numerical stability problems to be overcome. 

Nowadays,
the word \emph{cellular automata} has enlarged its meaning, including
any system whose elements do not move (in opposition to
``agent-based modeling''). Therefore, we now have cellular automata
on non-regular lattices, non-homogeneous, with probabilistic
dynamics (see \secref{DP}), etc.~(see for instance
Ref.\cite{ACRI2006}). They are
therefore considered more as a
``philosophy''
of modeling rather than a single tool. In some sense, cellular
automata (and agent-based) modeling is opposed to the spirit of
describing a phenomena using differential equations (or partial
differential equations). One of the reasons is that the language of
``automata'' is simpler and requires less training than that of
differential equations. Another reason is that at the very end, any
``reasonable'' problem has to be investigated using computers, and
while the implementation of cellular automata is straightforward
(even if careful planning may speed-up dramatically the simulation),
the computation of partially differential
equations is an art in itself.

\Subsection[DP]{Probabilistic Cellular Automata}

In deterministic automata, given a local configuration, the future
state of a cell is univocally determined. But let us consider the
case of measuring experimentally some pattern and trying to analyze
it in terms of cellular automata.
In time-series analysis,
it is common to perform averages over spatial patches and temporal
intervals and to discretize the resulting value. For instance, this
is the natural result of using a camera to record the temporal
evolution of an extended system, for instance the turbulent and
linear regions of a fluid. The resulting pattern symbolically
represents the dynamics of the original system, and if it is
possible to extract  a ``rule'' out of  this pattern, it would be
extremely interesting for the construction of a model. In general,
however, one observes that sometimes a local configuration is
followed by a symbol, and sometime the same local configuration
is followed by another one. One should conclude that the neighborhood
(the local configuration) does not univocally determines the
following symbol. 

One can extend the ``range'' of the
rule, adding more neighbors farther in space and
time~\cite{HiddenMarkovModel}. By doing so,
the ``conflicts''
generally reduce, but at the price of increasing the complexity of
the rule. At the extremum, one could have an automaton with infinite
``memory'' in time and space, that perfectly reproduces the observed
patterns but with almost none predictive power (since it is extremely
unlucky that the same \emph{huge} local configuration is encountered
again.

So, one may prefer to limit the neighborhood to some finite
extension, and accept that the rule sometimes ``outputs'' a symbol and
sometimes another one. One defines a local transition probability
$\tau(x_{\vec{i}}(t+1)| \vec{X}_{\vec{i}}(t))$ of obtaining a certain
symbol $x_{\vec{i}}$ at time $t+1$ given a local configuration 
$\vec{X}_{\vec{i}}$ at time $t$. Deterministic cellular automata
correspond to the case $\tau=0,1$. The parallel version of the
Lenz-Ising model can be re-interpreted as a probabilistic
cellular automaton.

Starting from the local transition probabilities, one can build up
the transition probability $T(\vec{x}|\vec{y})$ of obtaining
a configuration $\vec{x}$ given a configuration
$\vec{y}$ ($T(x|y)$ is the product of the local transition
probabilities $\tau$, for each site~\cite{Bagnoli:CellularAutomata}).
One can read the configurations $\vec{x}$ and $\vec{y}$ as indexes, so
that $T$ can be considered as a matrix. The normalization of
probability corresponds to the constraint
$\sum_{\vec{x}}T(\vec{x}|\vec{y}) = 1$, $\forall \vec{y}$.

Denoting with $P(\vec{x}, t)$
the probability of observing a given configuration $\vec{x}$ at time
$t$, and with $\vec{P}(t)$ the whole distribution at time $t$, we have
for the evolution of the distribution
\[
  \vec{P}(t+1) = T \vec{P}(t), 
\]
with the usual rules for the product of matrices and vectors.
Therefore, the transition matrix $T$ defines a Markov process, and
the asymptotic state of the system is given by the eigenvalues of
$T$. The largest eigenvalue is always 1, due to the normalization of
the probability distribution, and the corresponding eigenvector is
the asymptotic distribution. The theory of Markov processes says that
if $T$ is irreducible, \ie, it cannot be rewritten (renumbering rows
and columns) as blocks of noninteracting subspaces, like
\[
  T = \left(\begin{array}{cc}A&0\\0&B      \end{array}\right),
\]
then the second eigenvalue is strictly less than one and the
asymptotic state is unique. In this case the second eigenvalue
determines the convergence time to the asymptotic state. For finite
matrices, if all elements are greater than zero (and therefore
strictly less than one), the matrix $T$ is irreducible. However, in
the limit of infinite size, the largest eigenvalue may become
degenerate, and therefore there are more than one asymptotic state.
This is the equivalent of a phase transition for Markov processes.

For the parallel Lenz-Ising model, the elements of the matrix
$T$ are given by the product of local transition rules of the
Monte-Carlo dynamics. They depend on the choice of the algorithm, but
essentially have the form of exponentials of the difference in
 energy, divided by the temperature. Although a definitive
proof is still missing, it is plausible that matrices with all
elements different from zero correspond to some equilibrium model,
whose transition rules can be derived from an energy
function~\cite{Georges}.

Since probabilistic cellular automata are defined in terms of the
transition probabilities, one is free to investigate models that go
beyond equilibrium. For instance, if some transition probability
takes the value zero or one, in the language of equilibrium system
this would correspond to some coupling (like the $J$ of the Lenz-Ising
model) that become infinite. This case is not so uncommon in modeling.
The inverse of a
transition probability correspond to the average waiting time for the
transition to occur in a continuous-time model (one may think to
chemical reactions). Some transitions may have a waiting time so long
with respect to the observation interval, to be practically
irreversible. Therefore, probabilistic cellular automata (alongside
other approaches like for instance annihilating random walks) allow
the exploration of out-of-equilibrium phenomena.

\begin{figure}
\begin{center}
\includegraphics[width=0.5\columnwidth]{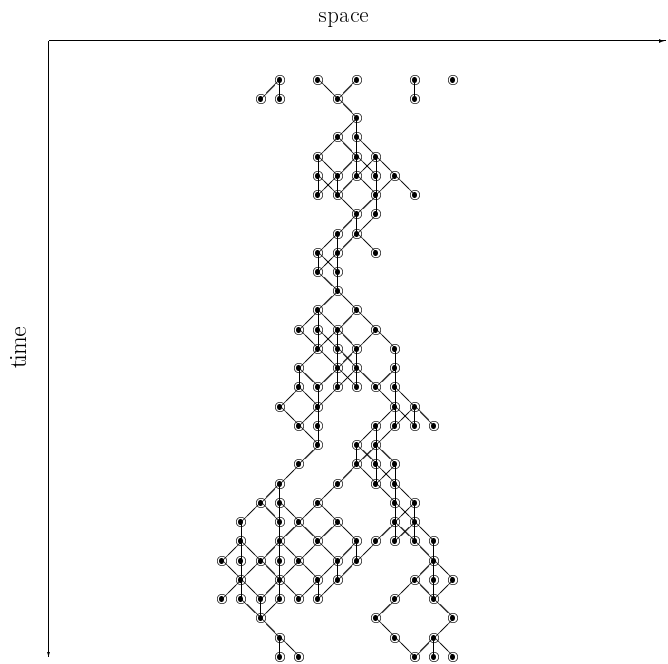}
\end{center}
\caption{\label{cluster} An example of a directed percolation
cluster.}
\end{figure}

One such phenomena is directed percolation, \ie, a
percolation process with a special direction (time)
along which links can only be crossed one-way~\cite{DP}. Let
think for instance to the spreading of an
infection in a one-dimensional lattice, with immediate recovery (SIS
model). An ill individual can infect one of both of his two neighbors
but returns to the susceptible state after one step. The paths of
infection (see Figure~\ref{cluster}) can wander in the space
directions, but are directed in the time directions. 

The parallel version of a directed percolation process can be mapped
onto probabilistic
cellular automata. The simplest case, in one spatial dimension 
and with just two neighbors, is called the
Domany-Kinzel model~\cite{DomanyKinzel}.  It is even more general than
the usual directed percolation, allowing ``non-linear'' interactions
among sites in the neighborhood (\eg, two wet sites may have less
probability of percolating than one alone). 

These processes are interesting because there is a 
\emph{absorbing
state}~\cite{Hinrichsen:SeveralAbsorbingStates,
Bagnoli:TwoAbsorbingStates}, which is the
dry state for the wetting phenomenon and the healthy state for the
spreading of epidemics. Once the system has entered this absorbing
state, it cannot exit, since the spontaneous appearing of a wet site
or of a ill individual is forbidden. For finite systems, the theory of
Markov chains says that the system will ``encounter'', sooner or
later, this state, that therefore corresponds to the unique asymptotic
state. For infinitely extended systems, a phase transition can occur,
for which wet sites percolate for all ``times'', the epidemics become
endemic, and so on.

Again these are examples of \emph{critical phenomena}, with exponents
different from the equilibrium case. 

\Subsection[CAAgents]{Cellular Automata and Agent-Based Simulations}

Cellular automata can be useful in modeling phenomena that can be
described in lattice terms. However, many phenomena requires ``moving
particles''. Examples may be chemical reactions, ecological
simulations, social models. When particles are required to obey
hydrodynamics constraints, \ie, to collide conserving mass,
momentum and energy, one can use lattice gas cellular automata or
the lattice Boltzmann equation. However, in general one is interested
in modeling just a macroscopic scale, assuming that what happens at
lower level is just ``noise''. According with the complexity (and
``intelligence'') assigned to particles, one can develop models based
on the concept of walkers, that move more or less randomly. From
the simulation point of view, walkers are not very different from the
graphs succinctly described above. In this case the identifier
$\vec{i}$ is just a label, that allows to access walker's data, among
which there are the coordinates of the walker (that may be
continuous), its status and so on.

In order to let walker interact, one is interested in finding
efficiently all walkers that are nearer than a given distance from
the one under investigation. This is the same problem one is faced
with when developing codes for molecular dynamics
simulations~\cite{MolecularDynamics}: scanning all walkers in
order to compute their mutual distance is a procedure that grows
quadratically with the number of walkers. One ``trick'' is that of
dividing the space in cells, \ie, to define an associated lattice.
Each cell contains a list of all walkers that are locate inside it. In
this way, one can directly access all walkers that are in the same or
neighboring cells of the one under investigation.

Moreover, one can exploit the presence of the lattice to implement on
it a ``cellular automaton'', that may interact with walkers, in order
to simulate the evolution of ``fields''. Just for example, the
simulation of a herd of herbivore that move according to the
exhaustion of the grass, and the parallel growing of the vegetation
can be modeled associating the ``grass'' to a cellular automaton, and
the herbivores to walkers. This simulation scheme is quite
flexible, allowing to implement random and deterministic
displacements of moving objects or agents, continuous or discrete
evolution of ``cellular objects'' and their
interactions. Many simulation tools tools and games are based on this
scheme~\cite{NetLogo,repast,SimCity},
and they generally allow the contemporary visualization of a
graphical representation of the state of the system. They are
valuable didactic tool, and may be used to ``experiment'' with these
artificial worlds. As often the case, the flexibility is paid in
terms of efficiency and speed of simulation. 

\section{Future Directions}

Complex systems, like for instance human societies, cannot be
simulated starting from ``physical laws''. This is sometimes
considered a weakness of the whole idea of studying such systems from
a quantitative point of view. We have tried to show that actually
even the ``hardest'' discipline, physics, always deals with models,
that have finally to be simulated on computers making various
assumptions and approximations. Theoretical physics is accustomed
since a long time to ``extreme simplifications'' of models, hoping to
enucleate the ``fundamental ingredients'' of a complex behavior. This
approach have proved to be quite rewarding for our understanding of
nature.

In recent years, physics have been seen studying many fields not
traditionally associated to physics: molecular biology, ethology,
evolution theory, neurosciences, psychology, sociology,
linguistics, and so on. Actually, the word ``physics'' may refer
either to the classical subjects of study (mainly atomic and
subatomic phenomena, structure of matter, cosmological and
astronomical topics), or to the ``spirit'' of the investigation, that
may apply to almost any discipline. This spirit is essentially that
of building simplified quantitative models, composed by many elements,
and study them with theoretical instruments (most of times,
applying some form of mean-field treatment) and with computer
simulations.

This approach has been fruitful in chemistry and molecular biology
and nowadays many physical journals have sections devoted to
``multidisciplinary studies''. The interesting thing is that not only
have physicists brought some ``mathematics'' into fields that are
traditionally more ``qualitative'' (which often corresponds to
``linear'' modeling, plus noise), but physicists have also discovered
many interesting questions to be investigated, and new models to be
studied. One example is given by the current investigations about
the structure of social networks, that were ``discovered'' by
physicists in the nontraditional field of social studies. 

Another contribution of physicists to this ``new way'' of performing
investigations, is the use of networked  computers. Since a long time,
physicists have used computers for performing computations, storing
data and diffusing information using the Internet. Actually, the
concept of what is now the World Wide Web was born at CERN, as a
method for sharing information among laboratories~\cite{Web}. The
high-energy physics experiments require a lot of simulations and data
processing, and physicists (among others) developed protocols 
to distribute this load on a grid of networked computers.
Nowadays, an European project aims to ``open'' grid
computing to other sciences~\cite{Grid}.

It is expected that this combination of quantitative modeling and
grid computing will stimulate innovative studies in many fields. Here
is a small sparse list of possible candidates:
\begin{itemize}
 \item Theory of evolution, especially for what concerns evolutionary
medicine.
\item Social epidemiology, coevolution of diseases and human
populations, interplay between sociology and epidemics.
\item Molecular biology and drug design, again driven by medical
applications.
\item Psychology and neural sciences, it is expected that the
``black box'' of traditional psychology and psychiatry will be
replaced by explicit models based on brain studies.
\item Industrial and material design.
\item Earth sciences, especially meteorology, vulcanology,
seismology.
 \item Archaeology: simulation of ancient societies,  reconstruction
of historical and pre-historical climates.
\end{itemize}

However, the final success of this approach is related to
the availability of high-quality experimental data that allow to
discriminate among the almost infinite number of models that can be
built.

\section{Glossary}
\begin{description}
\item[Nonlinear system:] A system composed by parts whose combined
effects are different from the sum of the effects of each part.
\item[Extended system:] A system composed by many parts connected by
a network of interactions that may be regular (lattice) or irregular
(graph).
\item[Graph, lattice, tree:] A graph is set of nodes connected by
links, oriented or not. If the graph is translationally invariant (it
looks the same when changing nodes), it is called a (regular) lattice.
A disordered lattice is a lattice with a fraction of removed links or
nodes. An ordered set of nodes connected by links is called a path. A
closed path not passing on the same links is a loop. A cluster is a
set of connected nodes. A graph can be composed by one cluster (a
connected graph) or more than one (a disconnected graph). A tree is a
connected graph without loops.
\item[Percolation:] The appearance of a
"giant component" (a cluster that contains essentially all nodes or
links) in a graph or a lattice, after adding or removing nodes or
links. Below the percolation threshold the graph is partitioned into
disconnected clusters, none of which contains a substantial fraction
of nodes or links, in the limit of infinite number of nodes/links.
\item[State of a system:] A complete characterization of a system at
a given time, assigning or measuring the positions, velocities and
other dynamical variables of all the elements (sometimes called a
configuration). For completely discrete systems (cellular
automata) of finite size, the state of the system is just a set of
integer numbers, and therefore the state space is numerable.
\item[Trajectory:] A sequence of states of a system,
labeled with the
time, \ie, a path in the state space.
\item[Probability distribution:] The probability of finding a system
in a given state, for all the possible states.
\item[Mean field:] An approximate technique for computing the value of
the observables of an extended system, neglecting correlations among
parts. If necessary, the dynamics is first approximated by a
stochastic process. In its simpler version, the probability of a state
of the system is approximated by the product of the probability of
each component, neglecting correlations. Since  the state of two
components that depend on a common ``ancestor'' (that interact with a
common node) is in general not uncorrelated, and this situation
corresponds to an interaction graph with loops, the simplest mean
field approximation consists in replacing the graph or the lattice of
interactions with a tree.
\item[Monte-Carlo:] A method for producing stochastic trajectories
in the state space designed in such a way that the time-averaged
probability distribution is the desired one.
\item[Critical phenomenon:] A condition for which an extended system
is correlated over extremely long distances.

\end{description}

\bibliographystyle{unsrt}
\bibliography{complexity}



\end{document}

%% file: macros.tex
\usepackage{ifthen}
\usepackage{xspace}

\newcommand{\eq}[2][]{%
	\ifthenelse{\equal{#1}{}}{%
		\[
		#2%
		\]%
	}{%
		\begin{equation}\label{Eq:#1}%
		#2%
		\end{equation}%
	}%
}
\newcommand{\meq}[2][]{%
	\ifthenelse{\equal{#1}{}}{%
		\begin{equation*}%
		\begin{split}%
		#2%
		\end{split}%
		\end{equation*}%
	}{%
		\begin{equation}\label{Eq:#1}%
		\begin{split}%
		#2%
		\end{split}%
		\end{equation}%
	}%
}

\renewcommand{\eqref}[1]{
	Eq.~(\ref{eq:#1})
}

\renewcommand{\vec}[1]{\boldsymbol{#1}}

\newcommand{\ie}{\emph{i.e.}\xspace}
\newcommand{\eg}{\emph{e.g.}\xspace}

\newcommand{\Section}[2][]{%
  \ifthenelse{\equal{#1}{}}{%
    \section{#2}%
  }{%
    \section{#2\label{sec:#1}}%
  }%
}

\newcommand{\Subsection}[2][]{%
  \ifthenelse{\equal{#1}{}}{%
    \subsection{#2}%
  }{%
    \subsection{#2\label{sec:#1}}%
  }%
}

\newcommand{\Paragraph}[1]{\paragraph{#1.}}

\newcommand{\secref}[1]{
  Sect.~\ref{sec:#1}%
}